\documentclass[a4paper]{article}

\usepackage{fullpage} 
\usepackage{pstricks}
\usepackage{multicol}
\usepackage{graphicx}
\usepackage{movie15}
\usepackage{hyperref}
\usepackage[version=3]{mhchem}
\usepackage{amsmath} 
\usepackage{amsfonts}
\usepackage{color}
\usepackage{soul}
\usepackage[margin=1.7cm,top=3cm,bottom=2.8cm,centering]{geometry}
\definecolor{rosepale}{rgb}{1.0, 0.7, 1.0}

\newcommand{\be}{\begin{equation}}
\newcommand{\ee}{\end{equation}}
\newcommand{\bea}{\begin{eqnarray}}
\newcommand{\eea}{\end{eqnarray}}

\title{Modeling generic aspects of ideal fibril formation\footnote{Reference: Michel, D. Modeling generic aspects of ideal fibril formation. 2016. J. Chem. Phys. 144, 035101}}

\author{Denis Michel \\
\\
      \begin{small} Universite de Rennes1-IRSET. Campus de Beaulieu Bat. 13. 35042 Rennes France. denis.michel@live.fr \end{small}
\\}

\date{} 

\begin{document}

\maketitle
\begin{multicols}{2}

\textbf{Abstract}. \ Many different proteins self-aggregate into insoluble fibrils growing apically by reversible addition of elementary building blocks. But beyond this common principle, the modalities of fibril formation are very disparate, with various intermediate forms which can be reshuffled by minor modifications of physico-chemical conditions or amino-acid sequences. To bypass this complexity, the multifaceted phenomenon of fibril formation is reduced here to its most elementary principles defined for a linear prototype of fibril. Selected generic features, including nucleation, elongation and conformational recruitment, are modeled using minimalist hypotheses and tools, by separating equilibrium from kinetic aspects and in vitro from in vivo conditions. These reductionist approaches allow to bring out known and new rudiments, including the kinetic and equilibrium effects of nucleation, the dual influence of elongation on nucleation, the kinetic limitations on nucleation and fibril numbers and the accumulation of complexes in vivo by rescue from degradation. Overlooked aspects of these processes are also pointed: the exponential distribution of fibril lengths can be recovered using various models because it is attributable to randomness only. It is also suggested that the same term "critical concentration" is used for different things, involved in either nucleation or elongation.

\section{Introduction}

Biomolecular complexes are generally made of a given number of components of various nature, so that their building is a finite process. By contrast, certain monotone fibrillar complexes are devoid of terminal capping moieties locking the polymerized components. As a consequence, their size is not precise but remains subject to depolymerization or growth depending on the concentration of their components and on the conditions. Understanding the fundamental mechanisms underlying this process, reviewed in \cite{Qian,Stryer,Firestone,Rangarajan,Andrews,Crespo,Ferrone,Cohen1,Cohen2,Cohen3,vanGestel}, is important as it can be physiological \cite{Pham}, of practical interest for food proteins \cite{Linden} and for biotechnological purposes in tissue engineering \cite{Ryan}, but also often pathological, for example in the cases of prion, amyloid or tau proteins in degenerative brains and mutant hemoglobin S in sickle cells. Although the most popular fibrillation substrates are proteins, fibrils can also be obtained with other molecules: short peptides such as polyalanines, polyglutamines, the hexapeptide VQIVYK from the tau protein \cite{Goux} or ultrashort peptides \cite{Hauser,Ryan}, and even non-protein small molecules \cite{Yang}, suggesting the existence of abiotic and perhaps prebiotic \cite{Cafferty} general fibrillation principles going beyond the sole field of proteins. Fibril formation can be studied kinetically \cite{Buell,Prigent,Schmit,Morris,Gillam} and at equilibrium \cite{Buell,Frieden,Qian}. Kinetic studies examine the transient behaviors of fibril formation, such as its sigmoidicity in time or the existence of lag-phases, whereas equilibrium corresponds to a stable state observed long after addition of a given amount of fibril components in a closed container. In vivo conditions do not correspond to equilibrium but are more complex, because of the presence of many additional molecular machineries and of the continuous synthesis and removal of fibril constituents. As fibril formation can be slow in vivo, it is often reproduced at higher rate in vitro, in optimized experimental conditions and using large concentrations of pure protein, either full length or reduced to short fibrillation-prone domains. \\
	The very rich literature in this field shows that fibril formation is a multifaceted phenomenon subject to interferences with a lot of parameters which cannot be taken into account simultaneously. It can proceed with different nucleation intermediates and different modes of polymerization \cite{Buell,Gillam,Kumar,Bellesia,Zhang} which depend on minor experimental changes or substitution of a single amino-acid, so that exhaustive models of fibril formation seem definitely unrealistic. In such a complex situation, it is often fruitful to recourse first to reductionist approaches for identifying general principles which can be combined and adapted to the specificities of real types of fibril. For comparison, ideal gases with monoatomic particles which obey perfectly elastic collision rules, do not exist in nature but have allowed to recover universal laws of statistical thermodynamics. This task would have been much more difficult by taking directly into account the precise shapes of the diatomic molecules of nitrogen and oxygen and the complex resulting collisions. In the present study, instead of starting from non-universal parameters such as the spatial organization of particular amyloid crystals, a series of basic features of fibril formation have been selected for focusing on thermodynamic principles. Only rudimental fibrils are considered, reduced to linear protofibrils before their possible subsequent association in twisted bundles. Their polymerization is analysed under its different facets, either direct or subsequent to a nucleation process, with or without random fragmentation, transient or in equilibrium, in vitro and in vivo. The appropriate tool for the discrete dynamic modelling of fibril polymerization is the master equation which allows to enumerate all the binding steps, such as for example in \cite{Cohen3}, but shortcuts are proposed here with continuous ordinary differential equations.
	
\section{Elongation: a question of binding propensity}

Let us call $ S $ a fibril-prone substrate. If a large initial concentration of monomers $ [S]_{0} $ is let in favorable physico-chemical conditions (temperature, pH, salts, etc.) for a very long time in a container of volume $ V $, $ S $ will be found in various forms including $ S_{1} $, the elementary subunit and $ S_{j} $, the fibrils containing $ j $ subunits. Fibril polymerization will be assumed noncooperative, which means that the same binding equilibrium constant $ K $ (the ratio between the association over dissociation rate constants $ K=k_{a}/k_{d} $), holds for all the steps. This approximation is acceptable as a change of $ K $ in the course of elongation would not seem in line with the monotonicity of long fibrils. In a closed system and in absence of cooperativity, the tendency to form fibrils is mainly dictated by two parameters: the affinity of the subunits for the tip of the fibril ($ K $) and the concentration of free subunits $ [S_{1}] $. The product of these two parameters gives a very useful dimensionless entity, written below $ K[S_{1}]=x $, which describes well the elongation propensity. Every polymerisation step $ S_{j}+S_{1} \rightleftharpoons  S_{j+1} $ is the resultant of the opposite reactions of association (rate $ k_{a} $) and dissociation (rate $ k_{d} $), which balance each other at equilibrium.

\begin{subequations} \label{E:gp}  
\begin{equation} k_{a}[S_{1}][S_{j}] = k_{d}[S_{j+1}] \end{equation} \label{E:gp1}
so that 
\begin{equation} x= K[S_{1}] = \dfrac{[S_{j+1}]}{[S_{j}]} \end{equation} \label{E:gp2}
\end{subequations} 
This simple expression suggests that as long as $ x>1 $, for every step $ j, \  [S_{j+1}]>[S_{j}] $ and fibrils are promoted. Conversely if the total concentration $ [S]_{0} $ is such that $ K[S]_{0}<1 $, no fibrils are expected. They can however form upon stochastic fluctuations and can eventually be stabilized by additional capping components. When $ [S_{1}] = 1/K =K_{d} $, for every $ j, \ [S_{j+1}]=[S_{j}] $, polymerization and depolymerization compensate each other. Hence, the so-called critical monomer concentration is close to $ K_{d} $ in supersaturated mixtures. The value of monomer concentration $ k_{\textup{off}}/k_{+} $ determined in \cite{Cohen3} using the master equation as the asymptotic limit of a dynamic system at infinite time, is completely equivalent to $ K_{d} $. When the initial input of monomers is much higher than $ K_{d} $ ($ [S]_{0} >> K_{d} $), calorimetry experiments showed that fibril elongation is an exothermic driven process \cite{Baldwin}. In this treatment, all the polymerization steps, including the first one, have the same binding constant, but in general, fibril formation can be restricted by a safety bolt: the difficulty to prime the process, called primary nucleation.

\section{Primary nucleation}
Adding a subunit to a nascent fibril is much more probable than the initial formation of a minimal fibril primer called nucleus. The nucleus is generally made of several monomers (Fig.1).

\begin{center}
\includegraphics[width=8.5cm]{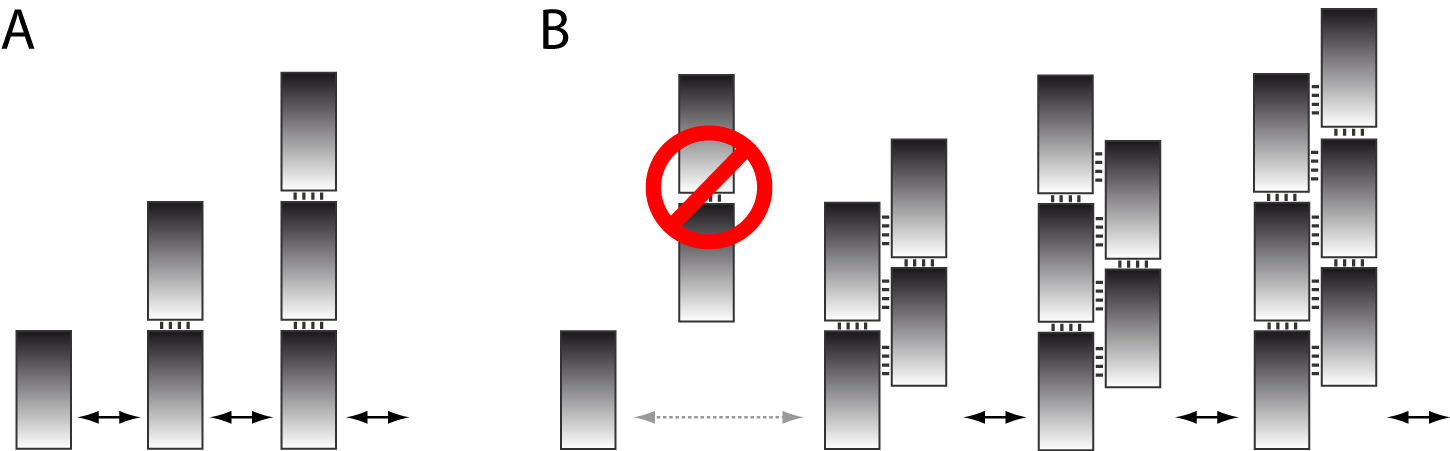} \\
\end{center}
\begin{small} \textbf{Figure 1.} $ n $-order reaction model to explain the restricting effect of primary nucleation on fibril formation. (A) Nucleation-independent and (B) nucleation-dependent fibril formation. Nucleation is necessary when the addition of one monomer necessitates interaction with several subunits from the nascent fibril. In the drawing (B), each subunit should interact with at least two orthogonal subunits to be stabilized. \end{small}\\

Real situations are however very complex because the nucleus components can be themselves pre-assembled composite building blocks, such as in the case of amyloid built by stacking layers of beta-sheets resulting from the assembly of beta-strands \cite{Kashchiev0}. \\
Several mechanisms have been proposed for modeling the limiting effect of nucleation. Three main characteristics of nuclei: (i) existence of lag time, (ii) critical concentration and (iii) seeding, have been simulated for amyloid fibrils using a lattice Monte Carlo procedure \cite{Zhang}. These properties can be also uncovered using classical mass action treatments when considering that nucleation results from either (i) a $ n $-order reaction, whose rarity results from the necessity of simultaneous collisions between several monomers \cite{Ferrone}, (ii) from a difficult dimerization \cite{Qian,Wegner,Congdon}, (iii) from a series of binding steps in which the tendency of dissociation is higher than that of association, as in the schemes of \cite{Hofrichter,Jarrett,Flyvbjerg}, or (iv) from the preliminary conformational activation of monomers, through spontaneous \cite{Chen} or induced transconformation \cite{Griffith,Prusiner} (see section 8.3).

\subsection{$ n $-order binding reaction}

The addition of a new subunit to a nascent fibril is stabilized by simultaneous interaction with several previously incorporated subunits (Fig.1B). As this is impossible for bimolecular reactions in the initial steps, polymerisation can start only from a nucleus. In the $ n $-order mechanism, the nucleus does not result from one-by-one addition of monomers but from a collision between more than two (say $ n $) subunits. This mechanism gives to the equilibrium binding constant a unit of concentration$ ^{1-n} $. When equilibrium is reached, this constant can be expressed with concentrations as 

\begin{equation} K_{n}^{*} = \dfrac{[S_{n}]}{[S_{1}]^{n}} \end{equation}
\noindent
where $ S_{n} $ is the primary nucleus \cite{Ferrone}. $ n $-order binding reactions are rare for spatio-temporal reasons. Trimolecular collisions are infrequent and, in addition, the correct orientation of more than two monomers for interacting properly is very improbable. This model of nucleation introduces a sort of cooperativity between the reactants (a genuine Hill cooperativity) which makes the formation of fibrils nonlinearly dependent on the dose of fibril components, as usually approximated \cite{Buell}. Higher than bimolecular reactions are so rare that they are generally neglected in biochemical modeling, but they can play a role in the particular case of the slow incubation of protein aggregation diseases, of which they are the limiting step. 

\subsection{Backward random walk}

The model of nucleation through a series of endothermic binding steps of \cite{Hofrichter,Jarrett,Flyvbjerg}, is more realistic than a global collision and is equivalent to a predominantly backward random walk whose frequency of completion is low. The mean time of completion of such walks dramatically increases with the number of steps, since it depends on the product of the ratio of the dissociation over the association rates for all the successive steps. \\
\\
These theoretical models of nucleation are of course simplified compared to real systems, such as cross-beta amyloid structures made of intertwined layers of beta-sheets. In the $ n $-order reaction and unfavorable random walk mechanisms, an ambiguity exists about the definition of the nucleus. In the simplified scheme of Fig.1B, the unstable version of the nucleus is a trimer, as generally assumed for actin filaments, whereas the minimal stable version of the nucleus is a tetramer. Indeed, the addition of the 4th monomer can also be considered as the first step of the elongation phase. In the rest of the present study, the nucleus will be considered as the greater unstable multimer before stabilization by elongation, in accordance with the view of \cite{Watzky,Kashchiev} considering the nucleus as the summit of unstability in the whole process. These unstable nuclei can then be locked by the ratchet effect of subsequent polymerization.

\section{Equilibrium behaviors of fibril formation}

\subsection{Detailed balance relations of elongation}
Equilibrium corresponds to in vitro systems in closed containers disconnected from the environment, but it is a macroscopic state that is not static at the microscopic level. Once formed, fibrils seem to be static structures, but in certain conditions, there are permanent exchanges between fibrils and free components \cite{Wetzel}, with a measurable "off rate" \cite{Gruning}. This reversible mode of complex formation can be briefly described in the case of uniform fibrils made of the same building block. Monomer addition will be supposed restricted to the apex of nascent fibrils, so that polymerisation is hierarchical, without need for complex statistical balancing between the microscopic and macroscopic equilibrium binding constants. The total amount of substrate $ [S]_{0} $ takes many forms whose concentrations obey

\begin{equation} [S]_{0} = [S_{1}]+2[S_{2}]+3[S_{3}]+.... =\sum_{j=1}^{V[S]_{0}}j[S_{j}] \end{equation}

These different concentrations are linked together by the equilibrium constant as follows

\begin{subequations} \label{E:gp}  
\begin{equation}[S_{2}]=K[S_{1}]^{2} \end{equation} \label{E:gp1}
\begin{equation}[S_{3}]=K[S_{1}][S_{2}]=K^{2}[S_{1}]^3 \end{equation} \label{E:gp2}
\noindent
etc,
\begin{equation}[S_{j}]=K^{j-1}[S_{1}]^{j} \end{equation}\label{E:gp3}
\end{subequations} 
\noindent
Hence, using $ x= K[S_{1}] $, Eq.(3) is close to

\begin{equation} K[S]_{0}=\sum_{j=1}^{\infty}jx^{j} \end{equation} 

When $ x<1 $, this series is convergent and its sum is
\begin{equation} K[S]_{0}=\dfrac{x}{(1-x)^2} \end{equation} 

The equilibrium concentration of fibrils containing $ j $ subunits is $ [S_{j}]= x^{j}/K $, where $ x $ is related to the total amount of protein through Eq.(6). For simplicity, fibrils are considered as molecular complexes with more than one subunit written $ [S_{F}]=\sum_{j=2}^{V[S]_{0}}j[S_{j}] $ so that $ [S]_{0} = [S_{1}]+[S_{F}] $. The concentration of fibrils is 
\begin{equation} [S_{F}] = \dfrac{x^{2}(2-x)}{K(1-x)^{2}} \end{equation} 
But in general, turbidity appears for fibrils larger than dimers, say $ r $ or more subunits. The concentration of turbid fibrils is
\begin{equation} \begin{split} [S]_{\geq r}& = \dfrac{x}{K}\sum_{j=r}^{\infty}jx^{j}=\dfrac{x}{K}\left (\sum_{j=0}^{\infty}jx^{j}-\sum_{j=0}^{r-1}jx^{j}  \right )\\ &= \dfrac{rx^{r}-(r-1)x^{r+1}}{K(1-x)^{2}} \sim \dfrac{rx^{r}}{K(1-x)} \end{split} \end{equation}

The combination of two parameters $ [S]_{0} $ and $ K $ dictates the fate of the system. Using Eq.(6), the final amount of non-monomeric substrate $ [S_{F}]= [S]_{0} - [S_{1}] $ can be defined as a function of the total amount of substrate $ [S]_{0} $ and of the binding equilibrium constant $ K $ as follows, 
\begin{equation} [S_{F}]=[S]_{0}-\dfrac{1}{K}\left (1+\dfrac{1}{2K[S]_{0}}-\sqrt{\dfrac{1}{K[S]_{0}}\left (1+\dfrac{1}{4K[S]_{0}}  \right )}\right ) \end{equation} 

whose shape is represented in Fig.2.

\subsection{Effect of primary nucleation on equilibrium fibrillation}

When the nucleation of a $ n $-mer nucleus is a prerequisite for fibril elongation, the total amount of fibril components spreads over the following species

\begin{equation} [S]_{0} = [S]_{1}+n[S_{n}]+(n+1)[S_{n+1}]+.... +(n+j)[S_{n+j}]+... \end{equation}

If nucleation proceeds through the $ n $-order mechanism postulated above, Eq.(10) can be rewritten

\begin{subequations} \label{E:gp}  
\begin{equation} \begin{split}  [S]_{0} = & [S_{1}]+K^{*}_{n}[S_{1}]^{n} (n+(n+1)x\\& +(n+2)x^{2}+....+(n+j)x^{j}+....) \end{split} \end{equation} \label{E:gp1}
with $ K^{*}_{n} $ defined in Eq.(2). This gives
\begin{equation} [S]_{0} = \dfrac{x}{K}+K^{*}_{n}\left (\dfrac{x}{K} \right )^{n} \left (\dfrac{n}{1-x}+ \dfrac{x}{(1-x)^{2}} \right )  \end{equation} \label{E:gp2}
\end{subequations} 
As previously in absence of nucleation, the total amount of fibrillar substrates can be defined. It depends now on three parameters: $ K, K^{*}_{n} $ and $ [S]_{0} $. $ [S_{F}] $ is a solution of the equation

\begin{equation}  \dfrac{n-(n-1)K([S]_{0}-[S_{F}])}{(1-K([S]_{0}-[S_{F}]))^{2}}K^{*}_{n}([S]_{0}-[S_{F}])^{n}-[S_{F}]=0 \end{equation}
where $ K $ and $ K^{*}_{n} $ are fixed constants depending on the molecular structures, but $ [S]_{0} $ can be modified. Fig.2 shows the concentration-dependence of fibrillation for given values of $ K $ and $ K^{*}_{n} $. The strong threshold effect obtained is a clear marker of nucleation. Using this criterion, the fibrillation of actin \cite{Oosawa-actin}, tau \cite{Congdon} and $ \beta $-lactoglobulin \cite{Linden}, is nucleation-dependent, whereas this is not the case, or very faintly, for collagen, based on the results of \cite{Na}. Concentration-dependent fibrillation curves without and with nucleation are clearly different at low substrate concentration (Fig.2) but converge towards the same asymptote $ [S]_{0}-1/K $ in supersaturated conditions.

\begin{center}
\includegraphics[width=7cm]{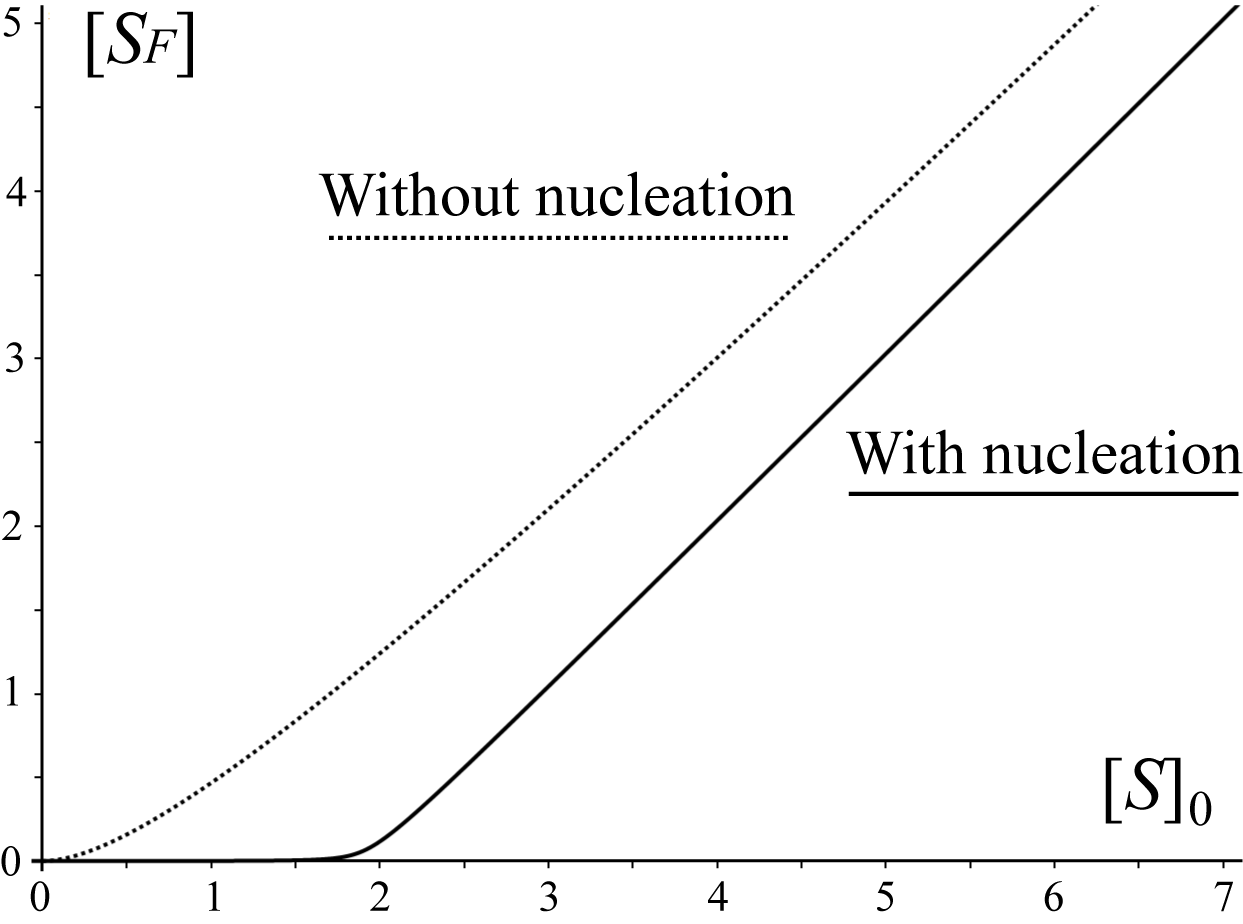} \\
\end{center}
\begin{small} \textbf{Figure 2.} Comparative concentration-dependence of fibrillation with and without nucleation. The amount of fibrillar substrate $ [S_{F}] $ is linearly dependent of the total amount of susbtrate $ [S]_{0} $, over a threshold concentration. The dotted curve obtained in absence of nucleation  is drawn to Eq.(9) with $ K= 0.5 $ c$ ^{-1} $. The plain line is obtained with primary nucleation by solving Eq.(12) for $ n=3 $ and then using the values $ K= 0.5 $ c$ ^{-1} $ and $ K^{*}_{n}=10^{-4} $ c$ ^{1-n} $, where c is the concentration unit. \end{small}\\

Hence, this treatment shows that the concept of critical concentration is ambiguous and used in the literature for different things. The concentration of substrate corresponding to the sharp bend of the curve with nucleation, is called critical concentration but has nothing to do with the critical concentration common to the two curves with and without nucleation (Fig.2). The first one is a function of nucleation and elongation parameters, while the second one is simply $ K_{d} $.

\subsection{Distribution of the subunits between the fibrils}

The concentration of fibrils of size $ j $ determined at equilibrium in \cite{Oosawa,Gillam,Cohen3}, is in the style

\begin{equation}[S_{j}]= K^{*}_{n}[S_{1}]^{n}(K[S_{1}])^{j-n} \end{equation}

This distribution can contain a peak when fibril growth and fragmentation are mixed in a master equation approach in \cite{Cohen3}. As $ [S_{1}] $ approaches $ 1/K $ near equilibrium in supersaturated conditions, all the fibril sizes would be equiprobable with Eq.(13), which does not predict the observed exponential length distribution. Alternative equations are proposed below. \\
In fact, the exponential distribution of fibril lengths, often verified experimentally, is the result of a series of interfering phenomena, including primary nucleation, complete fibril dissolution, fibril fragmentation and rejoining, which can not be taken into account simultaneously in modeling. But it is suggested here that this is not really a problem because even when considered by subsets, these mechanisms result in an exponential distribution typical of random events. Different mechanisms leading to this distribution will be proposed for fibrils. The first one is the distribution of subunits on nuclei. When reaching equilibrium with a supersaturated mixture ($ x \sim 1 $), $ [S_{1}]_{\textup{eq}} \sim K_{d} $ and $ [S_{F}]_{\textup{eq}} \sim [S]_{0}- K_{d} $ and fibril polymerization becomes a symmetric random walk because the tendencies of association and dissociation equalize. In the absence of fragmentation and in case of polarized (single end) elongation, the concentration of fibrils of any size corresponds to that of primary nuclei. Supposing that nuclei are stable and that their size is negligible compared to that of fibrils, the concentration of fibrils containing $ j $ subunits is

\begin{equation} [S_{j}]= [S_{F}] P(X=j) \end{equation} 
where $ P(X=j) $ is the probability that a given fibril contains $ j $ subunits. The mean length of fibrils is dictated by the concentration of nuclei $ [N] $. The number of nuclei is $ V[N] $ and the total number of subunits inserted in fibrils is $ V([S]_{0}-[S_{1}]) $. When equilibrium is reached, $ [S_{1}] $ fluctuates around $ K_{d} $ and the subunits freely and randomly jump between fibrils. These indiscernible subunits distribute over the nuclei in a manner similar to the random distribution of Maxwell-Boltzmann such that

\begin{subequations} \label{E:gp}
\begin{equation} P(X=j)=\dfrac{\textup{\large{e}}^{-j/\lambda }}{\sum_{k=0}^{V[S_{F}]}\textup{\large{e}}^{-k/\lambda }} \end{equation}  \label{E:gp1}
where $ \lambda $ is the mean number of subunits per fibril
\begin{equation}  \lambda =\dfrac{[S_{F}]}{[N]} \sim \dfrac{[S]_{0}-K_{d}}{[N]} \end{equation} \label{E:gp2}
\end{subequations} 

Hence, for a very large total number of subunits, the above equations yield

\begin{equation} [S_{j}]=([S]_{0}-K_{d})\left (1-\textup{\large{e}}^{-\frac{[N]}{[S]_{0}-K_{d}}} \right )\textup{\large{e}}^{-j \frac{[N]}{[S]_{0}-K_{d}}} \end{equation}
\noindent
whose shape is represented in Fig.3. 

\begin{center}
\includegraphics[width=7.5cm]{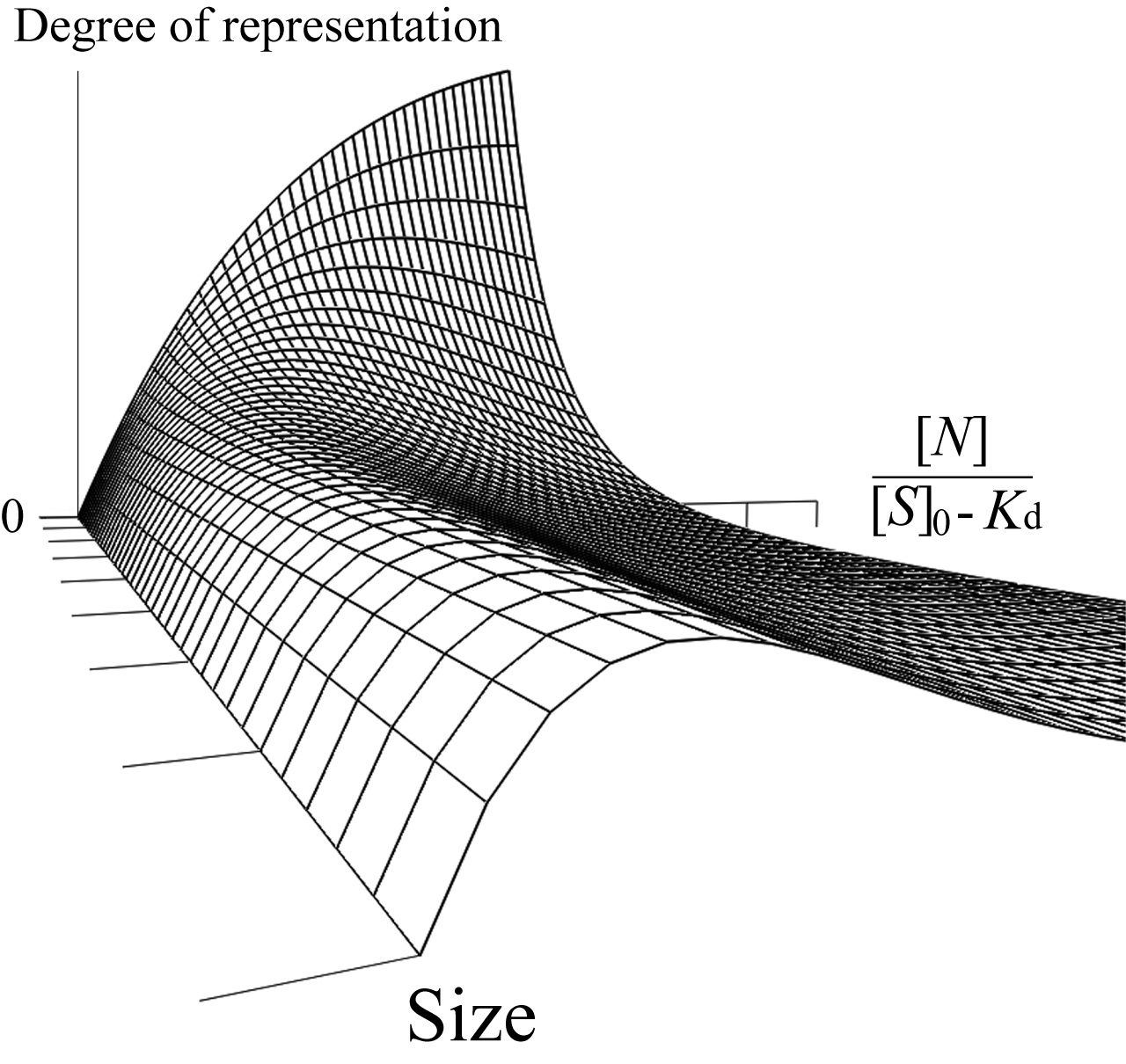} \\
\end{center}
\begin{small} \textbf{Figure 3.} Relative representation of the different sizes of fibrils according to Eq.(16). This curve shows that: (i) the sizes of the fibrils are exponentially distributed for a given ratio of concentration nuclei/substrates, (ii) fibrils are shorter when the concentration of nuclei is high compared to that of substrates, and (iii) fibrils are very long when the concentration of nuclei is very low (but not zero). \end{small}\\

After these equilibrium considerations, let us now summarize different kinetic properties of fibril formation.

\section{The transient primary nucleation}

The delay introduced by nucleation, combined to saturation, is a well established recipe for generating the sigmoidal kinetics of fibril accumulation according to the simplified system  described by Eq.(17). This system assumes the absence of fragmentation, nucleation through a $ n $-order reaction and irreversible elongation. The rates of nucleation, de-nucleation and elongation are written $ u, d $ and $ p $ respectively. Starting from monomers only, the concentrations of monomers ($ S_{1} $), of primary nuclei ($ PN $), of fibrils ($ F $) and of fibrillar substrate ($ S_{F} $), evolve according to

\begin{subequations} \label{E:gp}
\begin{equation} \dfrac{d[S_{1}]}{dt}=-nu[S_{1}]^{n}+nd[PN]-p[S_{1}]([PN]+[F]), \end{equation} \label{E:gp1}
\begin{equation} \dfrac{d[PN]}{dt}= u[S_{1}]^{n}-d[PN] -p[S_{1}][PN], \end{equation}\label{E:gp2}
\begin{equation} \dfrac{d[F]}{dt}= p[S_{1}][PN], \end{equation}\label{E:gp3}
\begin{equation} [S_{F}]= [S]_{0}-[S_{1}]-n[PN]. \end{equation}\label{E:gp4}
\end{subequations} 

In this system, the accumulation of primary nuclei is transient (Fig.4), because fibril elongation progressively depletes the medium in free monomers thereby forbidding further formation of primary nuclei.  

\begin{center}
\includegraphics[width=8cm]{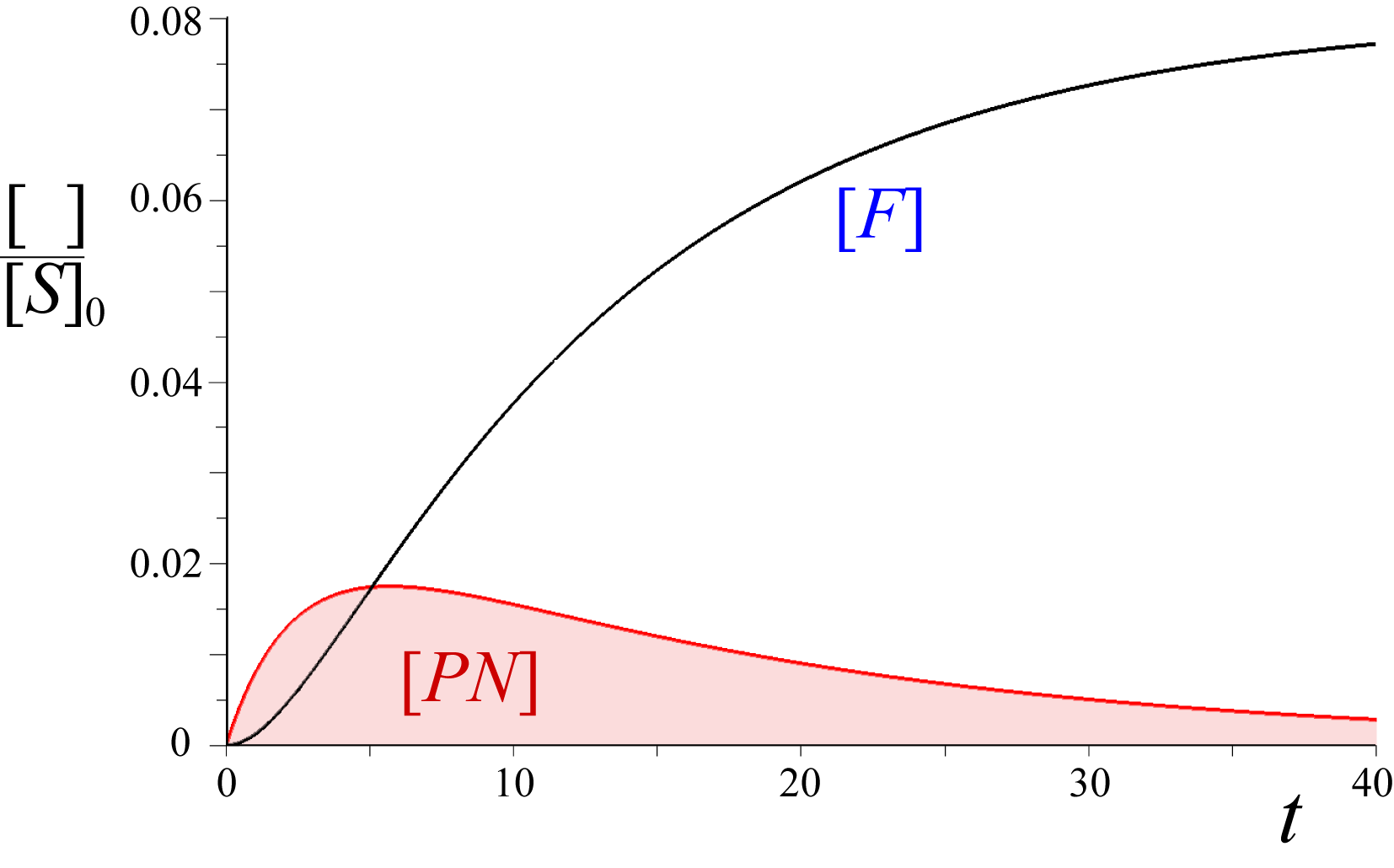} \\
\end{center}
\begin{small} \textbf{Figure 4.} Transient rise of free primary nuclei according to Eq.(17) with $ u=0.1 $ c$ ^{1-n} $ t$ ^{-1}$, $ d=1 $ t$ ^{-1}$, $ p=3 $ c$ ^{-1} $ t$ ^{-1}$, and $ n=3 $, where c is a concentration unit and t is a time unit adapted to the kinetic specificity of the system. \end{small}\\

\section{Kinetics of elongation}

\subsection{Stepwise kinetics}

Even in the absence of nucleation and saturation, the mere process of stepwise monomer addition is expected to give a sigmoidal increase of turbidity, more or less fugitive depending on the polymerization rate. For example, let us suppose that fibril formation is measured by turbidimetry and that turbidity is generated by fibrils containing $ r $ subunits or more, appearing when $ [S_{1}] $ remains much higher than the concentration of nuclei and the pseudo-first order monomer addition rate $ \beta = p[S_{1}] $ is roughly constant. The concentration of the substrates included in turbid fibrils is

\begin{subequations} \label{E:gp}
\begin{equation} [S]_{\geq r}\approx [S]_{0} P(X \geq r)\end{equation} 
where $ P(X \geq r) $ is the probability that a given fibril includes $ r $
or more subunits, this probability is the complement to unity of the sum of the probabilities to contain 0, 1, 2, 3,.... until ($ r-1 $) subunits
\begin{equation}  P(X \geq r)=1-\sum_{j=0}^{r-1}P(X=j) \end{equation}\label{E:gp2}
where each $ P(X=j) $ is part of a system
\begin{equation} \dfrac{dP(X=j)}{dt}=\beta \left [P(X=j-1)-P(X=j)  \right ] \end{equation}\label{E:gp3}
which gives the series 
\begin{equation} [S]_{\geq r}(t) \approx [S]_{0}\left (1-\textup{\large{e}}^{-\beta t}\sum_{j=0}^{r-1}\frac{(\beta t)^{j}}{j!}  \right ) \end{equation}\label{E:gp4}
\end{subequations} 

giving a nice sigmoidal arrival time.

\subsection{Elongation freezes primary nucleation}

The fibril length distribution given by Eq.(16) is valid but not useful in this form because the concentration of nuclei is unknown. This number depends on the mechanism of nucleation, primary or secondary. In the absence of secondary nucleation, the number of primary nuclei is a "kinetic legacy" which depends on the delicate balance between nucleation and polymerization. Indeed, once the first nuclei appear, they prime polymerization, which in turn lowers $ [S_{1}] $, thereby limiting subsequent nucleation because it is critically dependent on a threshold of $ [S_{1}] $. Hence, the concentration of primary nuclei does not obey a simple rule because they can form only at a stage preceding the elongation phase during which $ [S_{1}] $ dramatically drops. The formation of free primary nuclei ($ PN $) is transient but in case of irreversible fibril formation, they become embedded in fibrils and prevented to dismantle. The final concentration of nuclei results from the accumulation of nuclei during the nucleation window

\begin{equation} [PN]_{\textup{final}}=\int_{t=0}^{\infty }[PN]_{(t)} \ dt \end{equation} 
where $ [PN]_{(t)} $ is a convergent function of time (Fig.4) governed by the system of Eq.(17). The locking of primary nuclei is ensured by polymerization which can be approximated as irreversible during the nucleation phase during which $ [S_{1}] $ is high. If polymerization is exothermic reversible and if $ [PN]_{\textup{final}} $ is small enough compared to that of the polymerized substrate, the number of fibrils is expected to remain stable. In this scenario, a past transient period of primary nucleation remains inscribed in the number of fibrils (in the absence of fragmentation). This puzzling kinetic legacy also holds for other phenomena such as the long-term maturation \cite{Ma}, slow gelation \cite{Buell,Buell2} and slow compaction of fibrils. In apparently similar conditions, old fibrils no longer dissolve upon dilution \cite{Schirmer}. This age-dependent apparent irreversibility can reflect slow inter-subunit induced fit mechanisms further stabilizing the fibrils or the formation of interlocked fibril bundles. These singular situations highlight the complexity of protein aggregation processes with their multiple time scale separations, making difficult to distinguish transient from equilibrium phenomena. A fundamental property of hierarchical exothermic polymerization is that the addition of every subunit prevents the depolymerization of previously incorporated subunits. When exothermic polymerisation follows endothermic nucleation, the initial polymerization steps work as a fluctuation ratchet stabilizing the unstable nucleus, as schematized in the simplified energy landscapes of Fig.5. 

\begin{center}
\includegraphics[width=6cm]{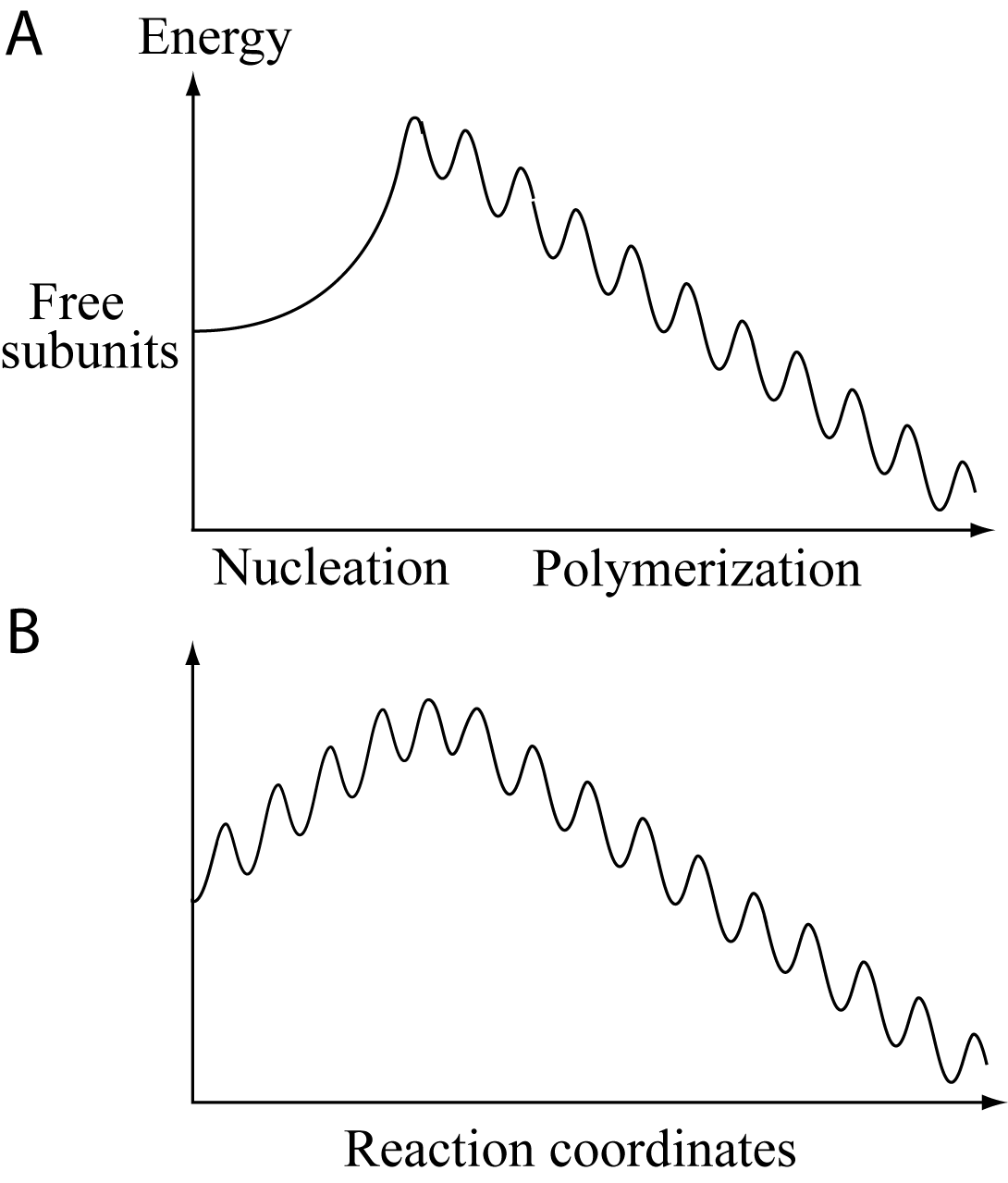} \\
\end{center}
\begin{small} \textbf{Figure 5.} Energy landscapes illustrating the ratchet effect of polymerization on nucleation, with two models: (A) One-step, $ n $-order nucleation reaction. (B) Nucleation made of a series of endothermic reactions. \end{small}\\

The concentration of primary nuclei established transiently can be frozen by (i) stabilization through the ratchet effect of elongation and (ii) by forbidding the formation of new primary nuclei because of the decrease of $ [S_{1}] $ (illustrated later in the example of Fig.10).

\section{Kinetics of fragmentation}

The need for nucleation can strongly limit the onset of fibrils, but once nascent fibrils are present, exothermic elongation can proceed at a relatively high rate. However, the overall elongation rate is constrained by the number of nuclei and fibril ends present in the mixture. In this respect, a particularly important parameter favouring fibril formation is the secondary formation of elongation primers. A predominant and natural way to achieve this end is the fragmentation of fibrils. Incidentally, this phenomenon is expected to be a major therapeutic problem for the development of fibril-dissolving drugs aimed at treating aggregation pathologies. The treatment can be worse than doing nothing when fibril fragmentation multiplies the number of elongation primers and of toxic oligomers. Fragmentation is expected to decrease in fibril systems at rest at a temperature where only small molecules can diffuse, but a permanent balance between fragmentation and dissolution/joining is however very likely in most real systems. 

\subsection{Elongation-fragmentation amplification cycles}
Fibril fragmentation can be spontaneous or provoked by certain proteins such as protein chaperones \cite{Winkler}. Fragmentation would strongly increase the overall rate of fibril formation through bypassing the need for rate limiting nucleation of monomers. This multiplication of elongation templates (seeding) exists naturally \cite{Tanaka} and can of course be intensified by experimental fragmentation \cite{Ohhashi,Xue}. If secondary nucleus doubling occurs over a critical size of fibrils, its modeling is equivalent to that of the well-known growth of a bacterial culture. In a closed container, as for bacterial growth curves, there is an exponential phase followed by a saturation phase.

\subsection{Random fragmentation}

The exponential fibril size distribution obtained in full equilibrium in Section 4.3, holds more generally, even in steady state for random modes of fragmentation. For example, let us postulate the existence, at the microscopic level, of invisible defective inter-subunit links randomly incorporated in the course of polymerization. These fragile points would break by Brownian resonance or unavoidable mechanical stress. If the mixture contains $ D $ defective links and $ S $ non-defective subunits, the probability that a given fibril contains $ \delta $ subunits is

\begin{subequations} \label{E:gp}
\begin{equation} P(L=\delta )=\left (\frac{S}{D+S}\right )^{\delta }\left (\frac{D}{D+S} \right ) \end{equation} \label{E:gp1}
whose continuous approximation is the probability density function of Boltzmann, like Eq.(15a). The probability that fibrils are longer than $ \delta $ is simply
\begin{equation} P(L>\delta )=\left (\frac{S}{D+S} \right )^{\delta } \end{equation}\label{E:gp2}
This equation can be rewritten using the mean number of subunits per fibril $ \lambda = S/D $
\begin{equation} P(L>\delta )=\left (\frac{\lambda }{1+\lambda } \right )^{\delta } \end{equation}\label{E:gp3}
For large $\lambda $ and $ \delta $, the continuous approximation of this function is
\begin{equation} P(L>\delta )=\textup{\large{e}}^{-\delta /\lambda } \end{equation}\label{E:gp3}
\end{subequations} 

Once again the geometric distribution is obtained. It can be truncated since small fibrils shorter than and around the nucleus size are absent, thereby generating a peak in the length distribution. The length distribution curves obtained by random fragmentation (this section) or random distribution over the primary nuclei without fragmentation (section 4.3), are closely resembling. Similar shapes of distribution curves have been obtained experimentally for the time-independent exponential distribution of Hemoglobin S fiber lengths \cite{Briehl} and for $ \alpha $-synuclein \cite{Raaij}.

\subsection{Fragmentation-driven fibril formation}

When $ S_{1} $ is abundant, one can define a pseudo-first order rate monomer addition to a fibril end $ \beta = p[S_{1}] $ where $ p $ is the intrinsic second order polymerization rate. For simplicity, the concentration of the monomers will be considered high enough to neglect the rate of depolymerisation. In the random mode of fibril fragmentation considered above, for a polarized growth and assuming that the rate of polymerization is constant, the doubling time (or fibril generation time), is $ \tau =  \lambda /\beta $ and as long as $ [S_{1}] $ is not limiting, the proliferation of fibrils follows

\begin{equation} [S_{F}](t)=[S_{F}]_{0} \ 2^{t/\tau } =[S_{F}]_{0} \ 2^{\beta t/\lambda } \end{equation}

\subsection{Saturation of fragmentation-dependent fibril formation}
\noindent
In a closed container not resupplied with monomers, the doubling time is not constant but evolves at each fibril generation such that 

\begin{subequations} \label{E:gp}
\begin{equation} [S_{F}](t+\tau (t))=2[S_{F}](t) \end{equation} \label{E:gp1}
and
\begin{equation}\tau (t)\geq \lambda /(p([S]_{0}-[S_{F}](t)) \end{equation}\label{E:gp2}
\end{subequations} 

Hence, $ [S_{F}] $ increases less and less rapidly because the doubling time is longer and longer as $ [S_{1}] $ decreases. The saturation thereby generated can be modeled simply. During the elongation phase driven by fragmentation, the role of primary nucleation is negligible and the time evolution of secondary nuclei $ [N] $ can be calculated as follows. The amount of substrate incorporated into fibrils satisfies

\begin{subequations} \label{E:gp}
\begin{equation} \dfrac{d[S_{F}]}{dt}= p[N][S_{1}]-q[N] \end{equation} \label{E:gp1}
where $ p $ and $ q $ are the rates of polymerization and depolymerization and with $ [S_{1}]=[S]_{0}-[S_{F}] $. If the nuclei duplicate after adding $ \lambda $ subunits per fibril, the concentration of elongation primers is related to that of polymerized substrate through $ [S_{F}]= \lambda [N] $. Hence, if calling $ a = (p[S]_{0}-q)/\lambda $, the previous equation becomes
\begin{equation} \dfrac{d[N]}{dt} = [N](a-p[N]) \end{equation}\label{E:gp2}
Starting at $ t_{0}=0 $ from a concentration of nuclei $ [N]_{0} < a/p $, we find
\begin{equation}  [N](t) = \dfrac{[N]_{0} \textup{\large{e}}^{at}}{1+\frac{p}{a}[N]_{0} (\textup{\large{e}}^{at}-1)}\end{equation} \label{E:gp3}
\end{subequations} 
\noindent
which gives a S-shaped growth curve starting from $ [N]_{0} $ (Fig.6). 

\begin{center}
\includegraphics[width=7.5cm]{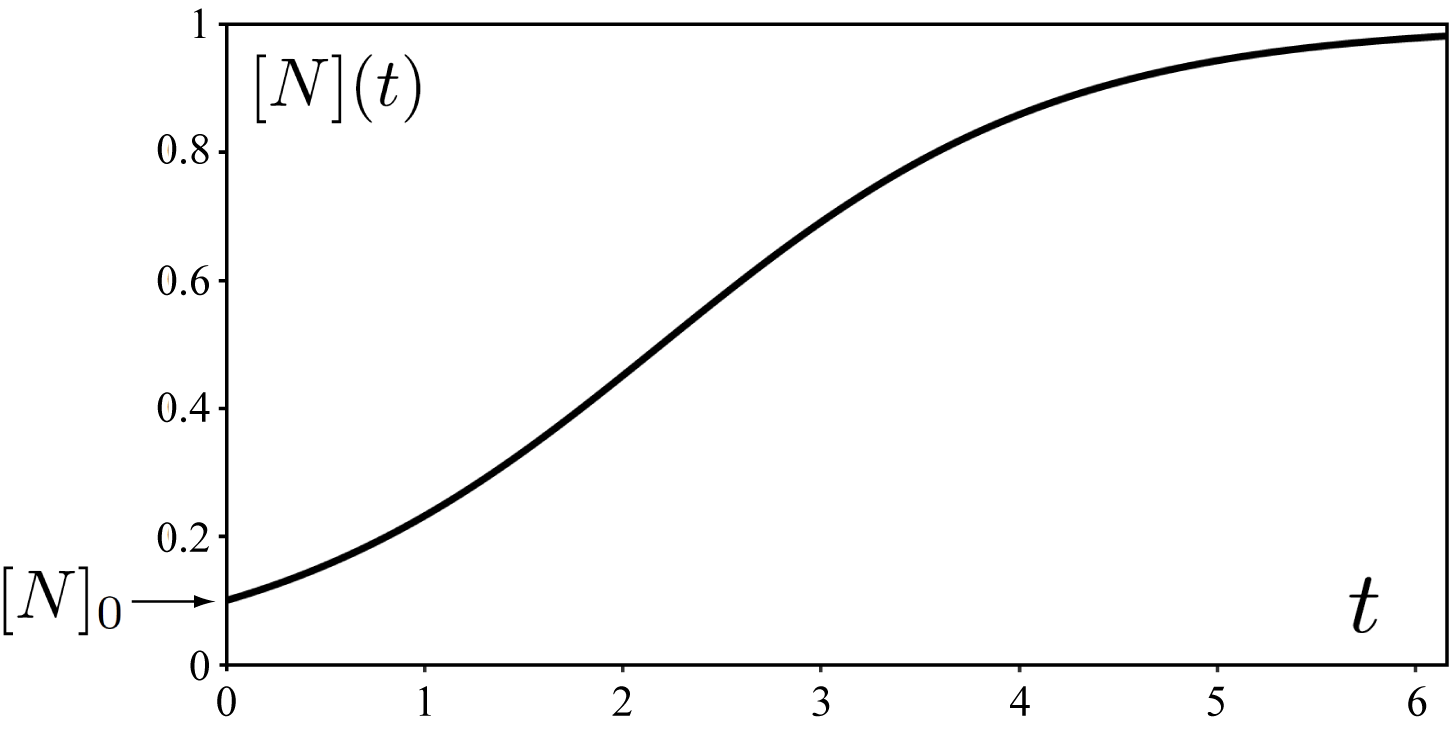} \\
\end{center}
\begin{small} \textbf{Figure 6.} Nucleus amplification and saturation. Scheme drawn to Eq.(23c) with $ [S]_{0}=30 $ c, $ [N]_{0} = 0.1 $ c, $ p=1 $ c$ ^{-1} $ t$ ^{-1}$, $ q=0 $ t$ ^{-1}$ and $ \lambda =40 $. \end{small}

\section{Kinetics of nucleation}

Nucleation is a rare event that is locked by the ratchet effect of subsequent polymerization. This situation can be modeled in different manners depending on the mechanism of nucleation retained, but with different equations, the $ n $-order reaction and the random walk models of nucleation yield very similar results, as described below.

\subsection{Nucleation based on a $ n $-order reaction}

In the scheme
\begin{center}
\ce{\textit{S}_{1}
<=>[\ce{\textit{u}}][\ce{\textit{d}}]
$\ce{\textit{S}_{n}}$
<=>[\ce{\textit{p}}][\ce{\textit{q}}]
$\ce{\textit{S}_{n+1}}$ }
\end{center}

$ u $ is a $ n $-order rate of one-step aggregation of $ n $ subunits into an extremely fragile nucleus $ S_{n} $, prompt to dismantle with a high rate $ d $, unless it is locked by polymerization (rate $ p $) supposed to be much higher than $ q $ during the nucleation-dominated phase when $ S_{1} $ is abundant. Starting from monomers only, the probability of first arrival to $ S_{n+1} $, is

\begin{subequations} \label{E:gp}
\begin{equation} P_{n+1}(t)=1-\textup{\large{e}}^{-\sigma t}\left (\frac{\sigma }{\mu }\sinh \mu t+\cosh \mu t  \right ) \end{equation} \label{E:gp1}
with
\begin{equation} \sigma = \left (d+p[S_{1}]+u[S_{1}]^{n}  \right )/2 \end{equation}\label{E:gp2}
and
\begin{equation} \mu = \sqrt{\sigma ^{2}-up[S_{1}]^{n+1}} \end{equation} \label{E:gp3}
\end{subequations} 
The appearance of primary nuclei is very transient and no longer sustainable when polymerization decreases $ [S_{1}] $ below the concentration threshold of nucleation. In this mechanism, time and substrate concentration work cooperatively. The priming of elongation is sharply dependent on the concentration of $ S_{1} $ and very rapid once this threshold is overstepped (Fig.7).

\begin{center}
\includegraphics[width=8cm]{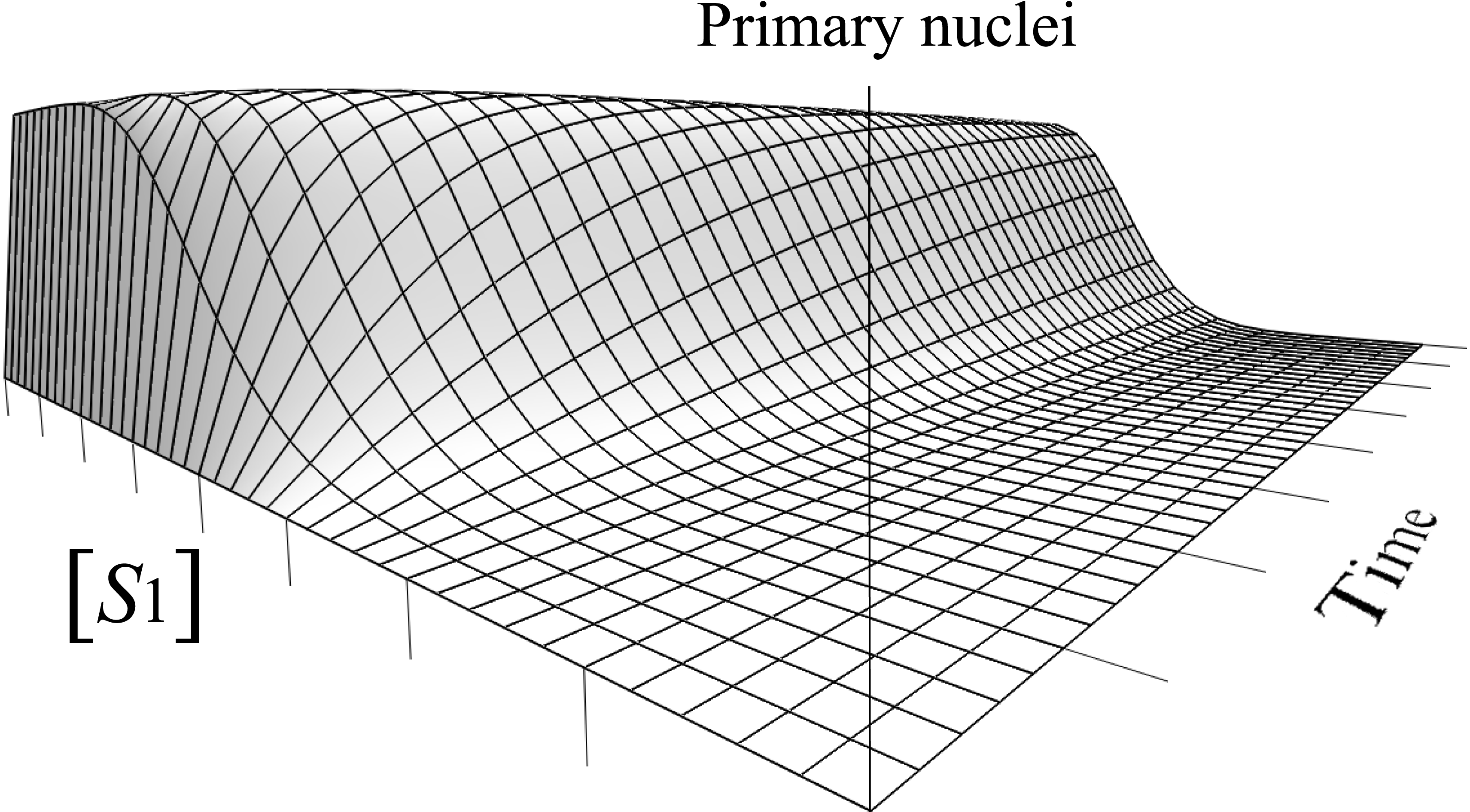} \\
\end{center}
\begin{small} \textbf{Figure 7.} Lag effect of $ n $-order nucleation on the kinetics of fibril formation. Curve drawn to Eq.(24) with arbitrary units ($ d=20 $ t$ ^{-1} $, $ p=20 $ c$ ^{-1} $ t$ ^{-1} $, $ u=0.1 $ c$ ^{1-n} $ t$ ^{-1} $ and $ n=6 $). No primary nuclei can form below a certain concentration of $ S_{1} $, which forbids the appearance of new primary nuclei after depletion of $ S_{1} $ caused by polymerization.
\end{small}\\

\subsection{Nucleation based on an endothermic random walk}

In this interesting model of nucleation, the components of the unstable nucleus $ S_{n} $ are not brought together in a unique collision, but pile up step by step ($ u $ in c$ ^{-1}$ t$ ^{-1}$ units) in a thermodynamically unfavorable manner \cite{Hofrichter,Jarrett,Flyvbjerg}. Nascent nuclei almost always dismantle before reaching the ($ n-1 $)th step. Considering that the nuclei can be stabilized by rapid elongation, the transition from $ n-1 $ to $  n $ is assumed irreversible, whereas all the previous transitions are endothermic. When the probability of backward jumps largely exceeds that of forward jumps ($ u[S_{1}]<<d $), the completion of the whole chain becomes similar to a single jump of very low probability. The mean time spent in step $ j $ compared to that spent in step $ n-1 $, is

\begin{equation} \dfrac{\left \langle t_{i}\right \rangle}{\left \langle t_{n-1}\right \rangle} = \dfrac{1-\left (\dfrac{d}{u[S_{1}]} \right )^{n-i}}{1-\dfrac{d}{u[S_{1}]}}\end{equation}

so that the partial nuclei are extremely rare and complete nuclei appear occasionally in a probabilistic manner, in a single reaction covering all the steps. This probability takes the form 

\begin{equation} P_{n}(t) \approx 1-\textup{\large{e}}^{-kt} \end{equation}

in which the global frequency $ k $ should now be defined. At each step $ j $, the probability to move forward instead of backward is $ u[S_{1}]/(d+u[S_{1}]) $, so that for the $ n-1 $ steps of a complete walk,
\begin{subequations} \label{E:gp}
\begin{equation} k=u[S_{1}]\left (\frac{u[S_{1}]}{d+u[S_{1}]}  \right )^{n-1} \end{equation} \label{E:gp1}
With the approximation $ d+u[S_{1}] \sim d $, one obtains
\begin{equation} k=\frac{(u[S_{1}])^{n}}{d^{n-1}} \end{equation}\label{E:gp2}
\end{subequations} 

and the time course of nuclei accumulation follows

\begin{equation} P_{n}(t) \approx 1-\textup{\large{e}}^{-((u[S_{1}])^{n}/d^{n-1})t}  \end{equation}

By itself, nucleation is not sigmoidal in time but the formation of fibrils becomes explosive only when $ P_{n}(t) $ is significant enough to prime elongation, thereby giving a sharp time course with a long lag phase. In the two models of nucleus formation ($ n $-order single reaction and $ n $-step random walk), the formation of primary nuclei is similarly dependent on the concentration of $ S_{1} $. This is not surprising given that a series of fast bimolecular reactions resembles a single collective reaction.

\subsection{Nucleation following recruitment of malconformed monomers}

In the model of nucleation inspired by prions, to be incorporated into fibrils, monomers should be first activated by transconformation. Prion proteins switch from a normal form $ S $, mainly alpha-helical, to a beta strand-rich form $ S^{*} $, called scrapie in reference to the corresponding pathology. In turn, newly converted monomers become monomer converters, thereby amplifying a recruitment chain. In the sporadic form of this pathology, not genetic (familial) or iatrogenic (contamination), the initial convertion is supposed to result from a stochastic, extremely improbable single event of very low mean rate $ k_{1} $, itself resulting from the conjunction of unknown microscopic events, making sporadic scrapie a "bad luck" disease. Once malconformed molecules $ S^{*} $ are present, they can then recruit their normal counterparts by inducing their transconformation, in an amplification chain long modeled in \cite{Watzky} and schematized in Fig.8. 
\begin{center}
\includegraphics[width=2cm]{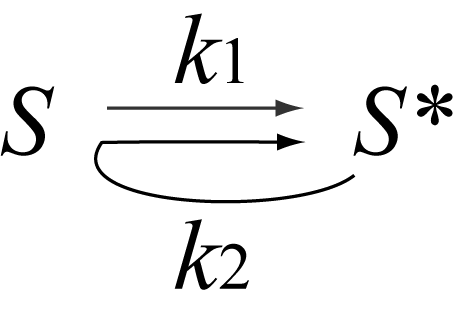} \\
\end{center}
\begin{small} \textbf{Figure 8.} Conformational contagion. Following a difficult initial conversion of $ S $ into $ S^{*} $, the converted moieties induce the conversion of their normal counterparts, in a snowball amplification chain. Three cases are possible in Scrapie with respect to the rate $ k_{1} $: (i) $ k_{1} $ is very low is sporadic cases, (ii) $ k_{1} $ is higher in genetic cases of familial mutations in the prion gene and (iii) $ k_{1} $ can remain very low in case of contamination by exogenous $ S^{*} $. \end{small}\\
\newline

Fig.8 can be translated into
\begin{subequations} \label{E:gp}
\begin{equation} \dfrac{d[S^{*}]}{dt}=k_{1}[S]+k_{2}[S^{*}][S]\end{equation} \label{E:gp1}
with, in a closed container,
\begin{equation} [S]=[S]_{0} - [S^{*}] \ \textup{and} \ [S^{*}](0)=0 \end{equation}\label{E:gp2}
The solution of this system, equivalent to that determined in \cite{Watzky,Crespo}, is
\begin{equation}[S^{*}](t)= \dfrac{k_{1}(\textup{\large{e}}^{(k_{1}+k_{2}[S]_{0})t}-1)}{k_{2}[S]_{0}+k_{1}\textup{\large{e}}^{(k_{1}+k_{2}[S]_{0})t}} \end{equation}\label{E:gp3}
\end{subequations} 

$ S^{*} $ accumulates in a sigmoidal manner similar to that of Fig.6, except that it starts from 0. The concentration of nuclei $ [N] $ can result form primary nucleation or breakage-mediated secondary nucleation etc. There is a multiplication of possible models by combining the different elementary submodels described previously. A minimalist scheme without fragmentation is represented in Fig.9. 

\begin{center}
\includegraphics[width=5cm]{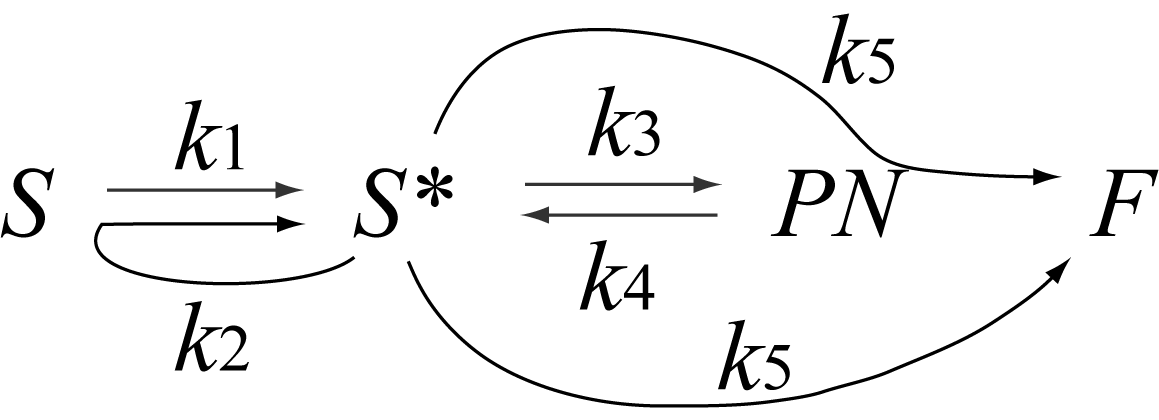} \\
\end{center}
\begin{small} \textbf{Figure 9.} Simplified model of protein aggregation following recruitment, completing that of Fig.8. $ S $ is the properly conformed substrate, $ S^{*} $ is the malconformed substrate and $ PN $ are the unstable primary nuclei. Contrary to nucleation that is considered as highly reversible ($ k_{4}>k_{3} $), elongation is assumed to be hard driven, with a quasi-irreversible transition $ k_{5} $. \end{small}\\

If the nucleus is not reduced to a monomer of $ S^{*} $, a second lag phase is generated by primary nucleation in addition to that caused by the inital conversion of $ S $ into $ S^{*} $ (Fig.10). A set of equations describing this situation is presented in Eq.(30) assuming the $ n $-order nucleation mechanism and no secondary nuclei. The time-dependent concentrations of the normal and malconformed substrate, primary nuclei and fibrillar substrate, evolve according to

\begin{subequations} \label{E:gp}
\begin{equation} \dfrac{d[S]}{dt}=-k_{1}[S]-k_{2}[S][S^{*}], \end{equation} \label{E:gp1}
\begin{equation} 
\begin{split}
\dfrac{d[S^{*}]}{dt}=& k_{1}[S]+k_{2}[S][S^{*}]-nk_{3}[S^{*}]^n \\
& +nk_{4}[PN]-k_{5}[S^{*}]([PN]+[F]),
\end{split}
\end{equation}\label{E:gp2}
\begin{equation} \dfrac{d[PN]}{dt}=k_{3}[S^{*}]^n-k_{4}[PN]-k_{5}[S^{*}][PN], \end{equation}\label{E:gp3}
\begin{equation} \dfrac{d[F]}{dt}=k_{5}[S^{*}][PN], \end{equation}\label{E:gp4}
\begin{equation} [S_{F}]=[S]_{0}-[S]-[S^{*}]-n[PN]. \end{equation}\label{E:gp5}
\end{subequations} 

Eq.(30) gives the sequential profiles shown in Fig.10, characterized by two lag phases. The fate of this system is very sensitive to the relative parameter values. For example, the recruiting influence of $ S^{*} $ can be moderated by its self-sequestration into fibrils, according to Eq.(30b). As illustrated in Fig.10B, in case of slow conformational recruitment, the competition between nucleation and recruitment can limit the total amount of malconformed substrate, and thereby of fibrils.

\begin{center}
\includegraphics[width=8cm]{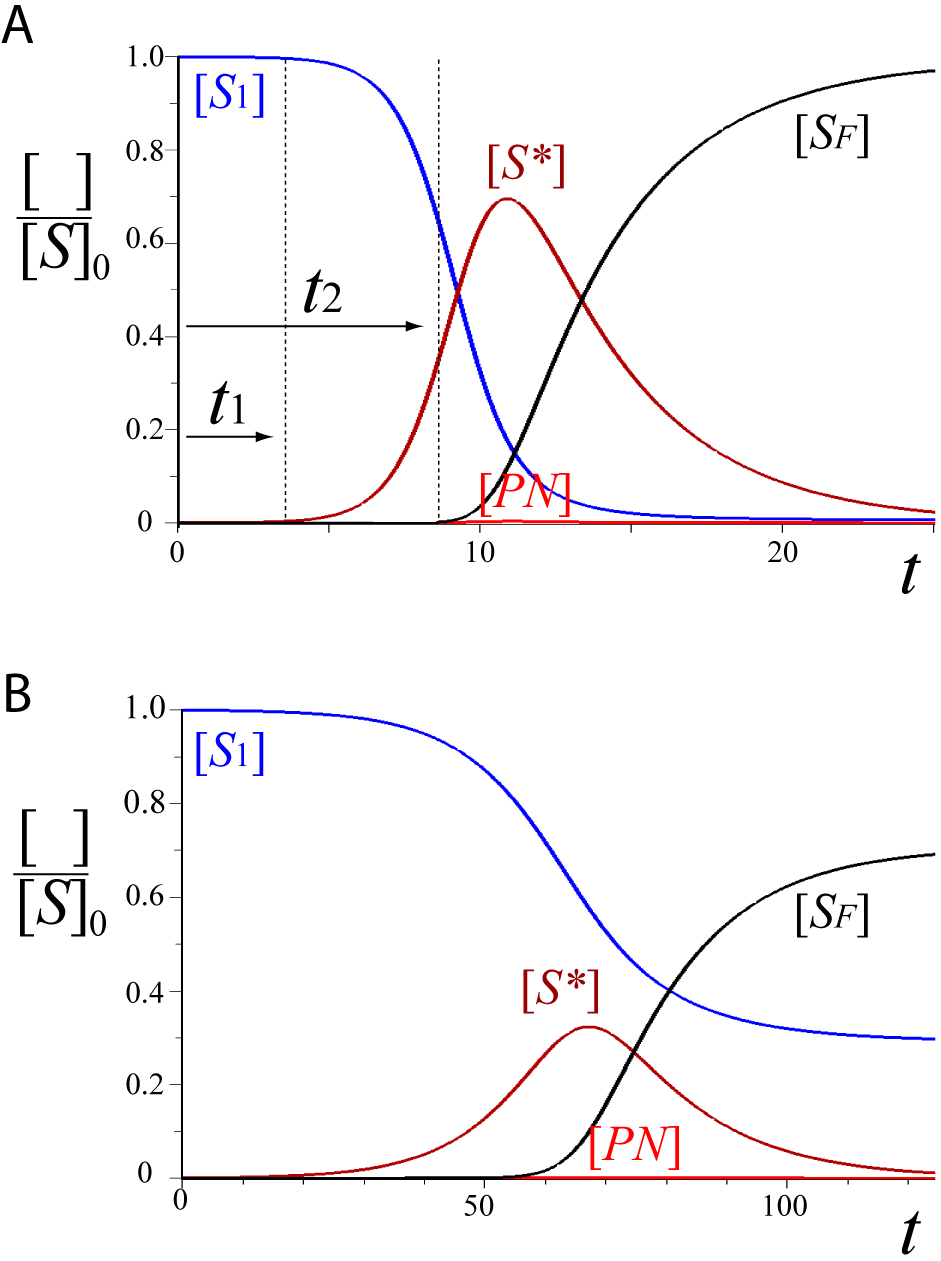} \\
\end{center}
\begin{small} \textbf{Figure 10.} Two lag phases in the onset of fibrils resulting from substrate transconformation. The first lag phase $ t_{1} $  corresponds to the priming of transconformation and the second one $ t_{2} $, is the formation of primary nuclei following accumulation of transconformed substrate above a concentration threshold. Curves drawn to Eq.(30) with arbitrary time units and parameters which are: (A) $ k_1=10^{-4} $ t$ ^{-1}$, $ k_2=1 $ c$ ^{-1}$ t$ ^{-1}$, $ k_3=0.06 $ c$ ^{1-n} $ t$ ^{-1}$, $ k_4=2 $ t$ ^{-1}$, $ k_5=6 $ c$ ^{-1} $ t$ ^{-1}$ and $ n=3 $). In (B) the only parameter changed is $ k_2=0.1 $ c$ ^{-1}$ t$ ^{-1}$. \end{small}\\

\section{Kinetic aspects of fibrils in vivo}
Most existing models start from a given amount of fibril constituents in a container, progressively incorporated into fibrils. In the previous example of prion contagion, an input of transconformational inducer is added at a given time point in the container. Living systems are however not closed containers but are by essence open. Fibrils ingredients can be synthesized and degraded, which seriously interferes with all the previous modeling attempts. Additional parameters should be introduced for in vivo modeling. Molecular crowding is important in vivo, particularly for nucleation, but it can be reproduced in vitro. In fact, the two most important parameters strictly specific of in vivo conditions are: (\textbf{i}) the relative kinetics of synthesis and fibril formation. If the synthesis of $ [S_{1}] $ is low, the exponential phase is very short or inexistent. (\textbf{ii}) The differential susceptibility to degradation of soluble and insoluble fibril constituents. This latter point is crucial in the development of slow conformational diseases. Indeed, it is classically assumed that protein aggregation results from a defect of protein degradation, but the inverse is even more important: an overlooked property of aggregates is their resistance to degradation, which can ensure physiological roles for programmed amyloid formation \cite{Pham}, but becomes harmful for amyloid diseases. 

\begin{center}
\includegraphics[width=4cm]{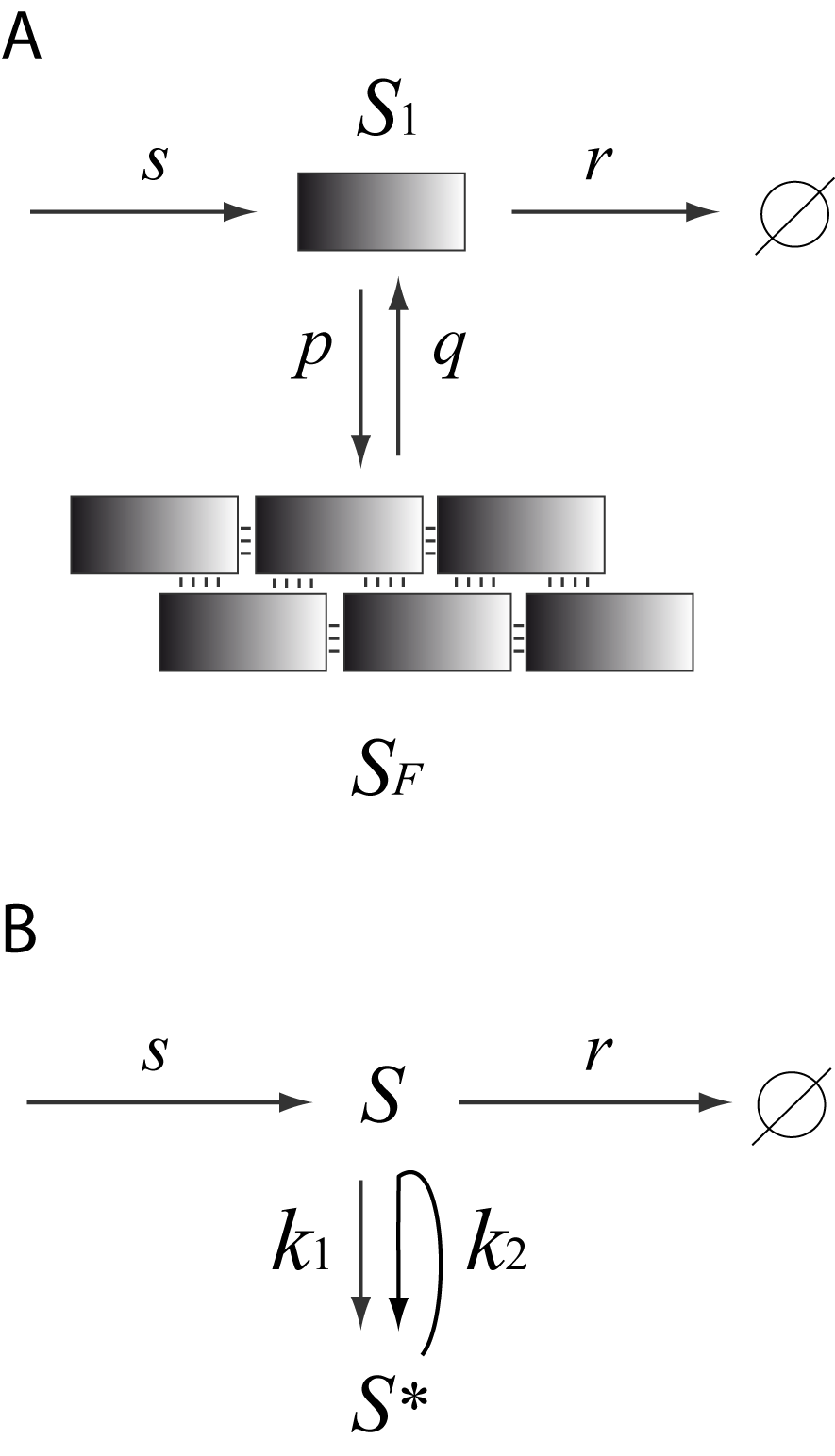} \\
\end{center}
\begin{small} \textbf{Figure 11.} Resistance to degradation of fibril constituents. (A) Protection by insolubilization following fibril polymerization. $ s $ and $ r $ are the rates of synthesis and removal of the soluble protein. $ p $ and $ q $ are the rates of polymerization and depolymerization. (B) Protection by transconformation into a protease-resistant structure. This scheme completes in vivo that of Fig.8 in vitro. \end{small}\\

The concentration of free fibril components is, as for all other soluble proteins, the steady state resultant of continuous synthesis and degradation. Fibril-prone proteins are synthesized as soluble and their rate of removal is generally a linear function of their soluble concentration. But once trapped into fibrils, they escape the degradation system according to the simplified scheme of Fig.11A. Starting from a concentration $ [S_{F}] $ of monomers polymerized into fibrils of mean size $ \lambda  $, the total concentration of substrate ($ [S]_{\textup{tot}} = [S_{F}]+[S_{1}] $) can increase unlimitedly in the cell according to the following system:

\begin{subequations} \label{E:gp}
\begin{equation} \dfrac{d[S]_{\textup{tot}}}{dt}=s-r([S]_{\textup{tot}}-[S_{F}]) \end{equation} \label{E:gp1}
\begin{equation} \dfrac{d[S_{F}]}{dt}= \dfrac{p}{\lambda }[S_{F}]([S]_{\textup{tot}}-[S_{F}]) \end{equation}\label{E:gp2}
\end{subequations} 

A limited rate of synthesis can restrict or forbid the existence of exponential phases, and the fibrils accumulate in a linear function of time (Fig.12). 

\begin{center}
\includegraphics[width=4cm]{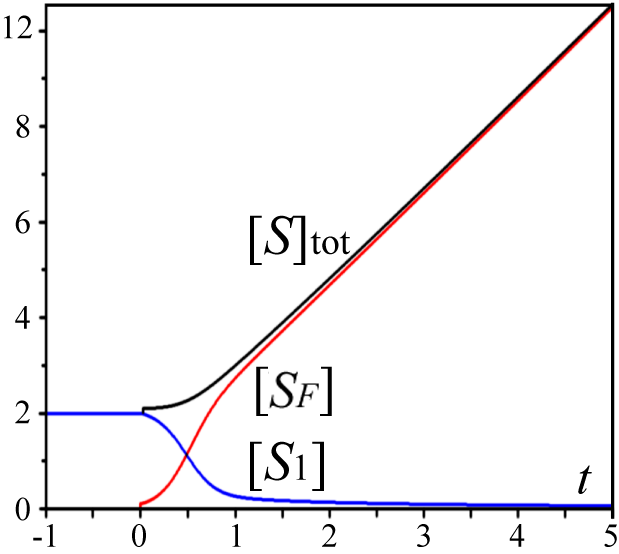} \\
\end{center}
\begin{small} \textbf{Figure 12.} Fibrils accumulate in vivo when the resistance to degradation provided by insolubilisation breaks the steady state. Schemes drawn to Eq.(31) with $ s=2$ c t$ ^{-1}$, $ p/\lambda =3$ c$ ^{-1} $ t$ ^{-1}$, $ r=1 $ t$ ^{-1}$, following addition of an input of fibrils $ [S_{F}](0)= 0.1 $ c, at time 0. \end{small}\\

The resistance to proteolysis has long been identified as the hallmark of scrapie \cite{McKinley} and the primary cause of prion accumulation after transconformation. Only the properly folded substrate can be submitted to normal synthesis/removal turnovers (Fig.11B). The formal dynamic treatment of the scheme of Fig.11B gives results like those shown in Fig.13. The comparison of panels C and D, in which $ S $ is continuously synthesised, is of particular interest. When only the soluble form is degradable (Fig.13C), $ S^{*} $ continuously accumulates. By contrast, when both $ S $ and $ S^{*} $ are submitted to the same degradation rate, one obtains a steady state (Fig.13D), in which $ s=r([S]+[S^{*}]) $. In this scenario, the fate of the disease depends on whether the value of $ [S^{*}] $ resulting from this system is lower or higher than the nucleation threshold, contrary to the case of Fig.13C where the threshold is necessarily reached sooner or later. In vivo modeling can be further complexified as desired, for example by introducing chaperones in the arena, but the general principle of an accumulation of $ S $ in the system through an escape from normal turnovers, remains essential.

\begin{center}
\includegraphics[width=8.5cm]{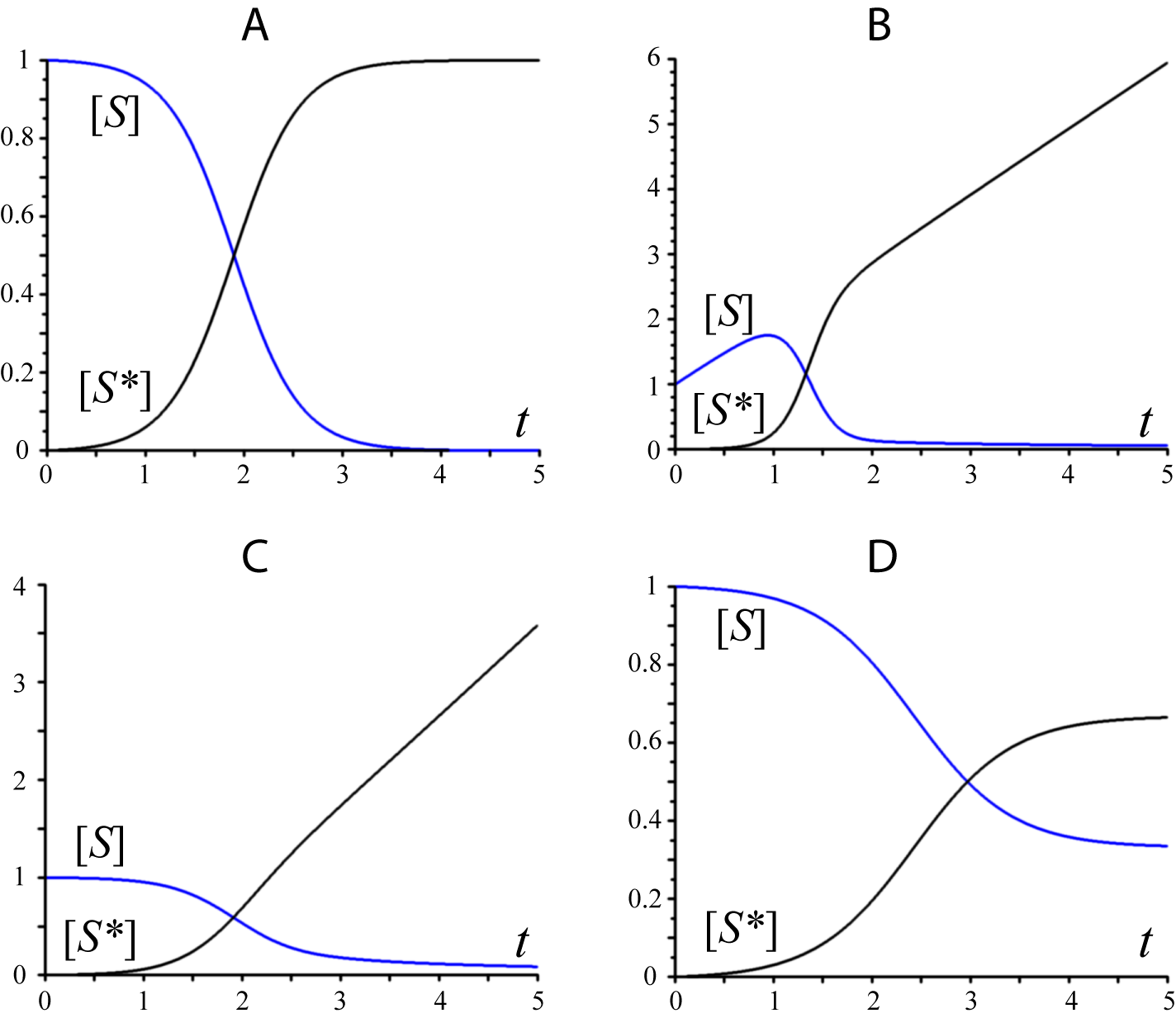} \\
\end{center}
\begin{small} \textbf{Figure 13.} Kinetics of prion transconversion based on the scheme of Fig.11B, in four cases: (A) No synthesis and no degradation of a finite initial amount of $ S $. (B) Synthesis, but not degradation of $ S $. (C) Synthesis and degradation of $ S $. (D) Synthesis of $ S $ and degradation of both $ S $ and $ S^{*} $. The rate constants used for these graphs are $ s=1 $ c t$ ^{-1}$, $ r=1 $ t$ ^{-1}$, $ k_{1} =0.01 $ t$ ^{-1}$ and $ k_{2} =3 $ c$^{-1} $ t$ ^{-1}$.  \end{small}\\

\section{Conclusions}
The polymorphic molecular process of fibril formation can be influenced by a virtually endless list of parameters, which depend on the specific cases and on the circumstances, such as protein disorganization, conformational changes, diffusion, viscosity, temperature, chaperones, pH, salts, metals, colloidal effects, secondary post-nucleation pathways, coexistence of multiples modes of nucleation, fibril branching and rejoining, and many other ones \cite{Morris,Gillam,Kumar}. To focus on the primary thermodynamic determinants of this complex biochemical process, a compendium of general rules is presented here, which highlights some problems in studying fibrils, such as (i) the difficulty to clearly separate transient from equilibrium phenomena, or (ii) to identify the precise causes of observations which can be explained by different models. \\
\indent
(i) In theory, regardless of how they have been established, equilibrium states can be entirely described using detailed balance rules. But it is very difficult in practice to determine when equilibrium is reached because of the coexistence of transient phenomena of very different time scales and of the long term imprinting of transient processes such as primary nucleation. \\
\indent
(ii) The sigmoidal kinetics of fibril growth sometimes considered as the hallmark of nucleation, can be underlied by different mechanisms. A good criterion for assessing the existence of limiting nucleation is a threshold effect at equilibrium (represented in Fig.2). This threshold is currently called a critical concentration but this not the same as the critical concentration of elongation. The distribution of fibril lengths has often been shown exponential \cite{Oosawa2,Oosawa,Morris,Raaij,Lee}. It is suggested that this distribution is a robust mark of randomness which holds for different models. All these resemblances could lead to misleadingly satisfactory curve fitting. \\
The main simplifying hypothesis facilitating the ordinary differential modeling used here, is the quasi-irreversibility of elongation. This approximation has the disadvantage of leading to the complete depletion of monomers but is quite realistic during nucleation and the transition phase between nucleation and elongation \cite{Powers}. These simplified tools allowed pointing original mechanisms involved in fibril formation: Elongation has ambiguous effects on nucleation: on the one hand stabilizing nuclei through a ratchet effect, and on the other hand, restricting the number of primary nuclei through lowering the concentration of monomers. In vivo, protein insolubilization is expected to play an essential role in amyloid accumulation through the escape from degradation.\\

\end{multicols}
\end{document}